\documentclass[journal, twoside, final]{IEEEtran}

\usepackage{cite}
\usepackage{graphicx}
\usepackage{amssymb, amsmath, amsthm}
\usepackage{amsfonts}
\usepackage{multirow}
\usepackage{mathrsfs}
\usepackage{color}
\usepackage{url}
\usepackage{flushend}
\usepackage{setspace}
\usepackage{placeins}
\usepackage{booktabs, multicol, multirow}
\allowdisplaybreaks

\newtheorem{theorem}{Theorem}
\newtheorem{remark}{Remark}

\begin{document}

\title{Fundamental Relations Between Reactive and Proactive Relay-Selection Strategies}

\author{Minghua~Xia,~\IEEEmembership{Member,~IEEE}, and Sonia~A\"{\i}ssa,~\IEEEmembership{Senior Member,~IEEE}
\thanks{
Manuscript received February  3, 2015; revised March 26, 2015; accepted March 28, 2015. This work was supported by a Discovery Grant from the Natural Sciences and Engineering Research Council (NSERC) of Canada. The associate editor coordinating the review of this manuscript and approving it for publication was O. A. Dobre.}
\thanks{
M. Xia is with Sun Yat-sen University, Guangzhou, 510275, China  (e-mail: xiamingh@mail.sysu.edu.cn). He was with the Institut National de la Recherche Scientifique (INRS),
University of Quebec, Montreal, QC, H5A 1K6, Canada.
}
\thanks{
S. A\"issa is with the Institut National de la Recherche Scientifique (INRS), University of Quebec, Montreal, QC, H5A 1K6, Canada (e-mail: aissa@emt.inrs.ca).
}
\thanks{
Digital Object Identifier 10.1109/LCOMM.2015.2418780}
}

\markboth{IEEE Communications Letters, accepted for publication} {Xia \MakeLowercase{\textit{et al.}}: Fundamental Relations Between Reactive and Proactive Relay-Selection Strategies}

\maketitle

\pubid{1558-2558\copyright~2015 IEEE. Personal use is permitted, but republication/redistribution requires IEEE permission.}

\pubidadjcol

\begin{abstract}
Two major relay-selection strategies widely applied in cooperative decode-and-forward (DF) relaying networks, namely, reactive relay selection (RRS) and proactive relay selection (PRS), are generally looked upon as independent and studied separately. In this paper, RRS and PRS are proven to be equivalent with respect to the end-to-end outage probability from the first principle, i.e. their respective relay-selection criteria. On the other hand, RRS is shown to be superior to PRS with respect to the end-to-end symbol error rate. Afterwards, a case study of a general DF relaying system, subject to co-channel interferences and additive white Gaussian noise at both the relaying nodes and the destination, is performed to explicitly illustrate the aforementioned outage equivalence. These fundamental relations provide intuitive yet insightful performance benchmarks for comparing various applications of these two relay-selection strategies.
\end{abstract}

\begin{IEEEkeywords}
\noindent Decode-and-forward (DF) relaying, proactive relay selection (PRS), reactive relay selection (RRS).
\end{IEEEkeywords}

\section{Introduction}
\IEEEPARstart{W}{ith} the technology of cooperative relaying, larger network coverage can be achieved by employing dual-hop transmissions, whereby the communication between a source and its far-end destination is performed over a single branch comprising an intermediate relay \cite{SuraweeraTWC0906}, or through multiple single-relay branches, where significant spatial diversity gain can be achieved \cite{ForghaniWCL1302}. In particular, the spatial diversity gain increases linearly with the number of relays. However, when multiple relays work simultaneously, they will interfere with each other if they operate within the same frequency band, or occupy more spectral bandwidth if they work in different band slots, needless to say that the synchronization among relays is challenging in practice. To resolve these issues, the technique of relay selection emerged as a practical solution, with no deterioration in the achievable diversity gain.

In the decode-and-forward (DF) relaying networks, there are two typical relay-selection strategies: reactive relay selection (RRS) and proactive relay selection (PRS) \cite{BletsasTWC0709}. For the RRS, the relaying nodes which can successfully decode the received signals in the first hop form a decoding set, from which the relay that delivers the highest signal-to-interference-and-noise ratio (SINR) at the destination is chosen as the best relay to re-encode and forward the source's signals to the destination, while the other relays remain silent. For the PRS strategy, on the other hand, the relay which maximizes the minimum SINRs at the first and second hops is chosen as the best relay. The latter strategy is also known as ``max-min" criterion, and is widely studied in the literature due to its ease of mathematical tractability \cite{KrikidisTWC0906}. Based on a simple dual-hop relaying model with the channel at each hop being subject to Rayleigh fading \cite[Eq. (4)]{BletsasTWC0709}, the end-to-end (e2e) outage probabilities pertaining to RRS and PRS were explicitly computed and they were shown to be identical \cite[Eqs. (15) and (20)]{BletsasTWC0709}. Despite the primitivity of this particular computation, it was not inferred that RRS and PRS are equivalent in general cases, because one cannot conclude universal relations from the observation of particular instances. To reveal their universal equivalence with respect to the e2e outage probability, preliminary inspection from the first principle, i.e. their respective relay-selection criteria, will be carried out in this paper.

Although the technique of relay selection benefits improving the performance of cooperative relaying networks, co-channel interference (CCI) originating from external interfering sources will significantly deteriorate the system performance. In order to explicitly reveal the effect of CCI on performance, many efforts were made by researchers in the community. For instance, the authors of \cite{ZhongTCOM1003, JuTVT1311} studied the effect of CCI at the destination, where the relaying nodes were assumed to be interference-free. In \cite{YangPIMRC09}, the CCI was considered at both the relays and the destination, but additive white Gaussian noise (AWGN) was neglected at any node in the network, which makes the corresponding results inaccurate in the low and medium SNR regions. Taking both CCI and AWGN into account at the relays and the destination, the system performance of DF relaying over Rayleigh fading was studied in \cite{GuCL1310}. Recently, the performance of optimum combining used in a DF relaying network operating in the presence of Nakagami fading and CCI at the relays and the destination was analyzed in \cite{SuraweeraWCL1310}.

\pubidadjcol

In this paper, after introducing the principles of the RRS and PRS strategies, we first prove their universal equivalence with respect to the e2e outage probability. Moreover, RRS is shown to be superior to PRS with respect to the e2e symbol error rate. Then, a case study of a general DF relaying system in the presence of multiple CCIs and AWGN at both the relaying nodes and the destination, is performed to further illustrate their outage equivalence. The newly revealed fundamental relations between RRS and PRS in terms of outage probability and symbol error rate provide intuitive yet insightful performance benchmarks for comparing various applications of these two relay-selection strategies.

\section{Principles of Reactive and Proactive Relay-Selection Strategies}
\label{sec:system_model}
We consider a multi-branch dual-hop DF relaying network, where the source, $\mathrm{S}$, sends data to its destination, $\mathrm{D}$, via $N$ independent branches, with branch $n$ comprising one intermediate relay, $\mathrm{R}_{n}$. Each relay as well as the destination is interfered by $L$ external co-channel interferers, $\mathrm{I}_l$, $l = 1, \cdots, L$.

As far as the RRS strategy is concerned, at the end of the first transmission phase, the relays successfully decoding the received signals form the decoding set, denoted $\mathcal{C}$ with the cardinality $|\mathcal{C}| = k$. Notice that {\it the decoding at relay $n$ is assumed to be successful if no outage event happens during the first transmission phase} \cite{BletsasTWC0709, LanemanTIT0412}. Then, in the second transmission phase, the relay belonging to the decoding set and having the highest SINR at the second hop is chosen as the best relay to re-encode and forward its received signal to the destination, while the other relays keep silent. Mathematically speaking, the index $n_1$ of the best relay, $R_{n_1}$, is chosen according to the criterion:
\begin{equation}
\label{Eq_RRS}
n_1 = \arg\max\limits_{r \in \mathcal{C}} {\Gamma_{r, \mathrm{D}}},
\end{equation}
where $\Gamma_{r, \mathrm{D}}$ refers to the received SINR corresponding to the second-hop link from relay $r$ to destination $\mathrm{D}$.

On the other hand, if the PRS strategy is adopted, the relay which maximizes the minimum between the SINRs received at the first and second hops is chosen to be the best relay. The chosen relay, after decoding the received source signal successfully, re-encodes it and forwards it to the destination. During these two consecutive transmission phases, the other relays always keep silent. Thus, the index of the best relay, denoted $n_2$, is determined as per the following rule:
\begin{equation}
\label{Eq_PRS}
n_2 = \arg\max\limits_{n=1, \cdots, N}\min\left\{\Gamma_{\mathrm{S}, \mathrm{R}_n}, \, \Gamma_{\mathrm{R}_n, \mathrm{D}}\right\},
\end{equation}
where $\Gamma_{\mathrm{S}, \mathrm{R}_n}$ and $\Gamma_{\mathrm{R}_n, \mathrm{D}}$ denote the received SINRs corresponding to the first-hop link ($\mathrm{S} \to \mathrm{R}_n$) and the second-hop link ($\mathrm{R}_n \to \mathrm{D}$), respectively.

Since the RRS strategy needs to transmit to all relays at the first hop whereas the PRS transmits only to the chosen best relay, it is clear that the latter decreases the requirement on spectral bandwidth if the relays work with different frequency bands, and improves the energy efficiency as only the best relay is active while the others remain idle. However, the RRS needs only local channel state information (CSI) at the second hop when performing relay selection, whereas the PRS requires global CSI (i.e. both the first and second hops) to make decision, which is hard to be collected in real-world networks with large number of branches.

In the next section, we investigate the relations between these two relay-selection strategies with respect to the e2e outage probability and symbol error rate.

\section{The Relations Between Reactive and Proactive Relay-Selection Strategies}
The outage equivalence between the RRS and PRS relay selection approaches is summarized in the following theorem.
\begin{theorem}
 \label{Theorem-1}
For cooperative DF relaying networks with $N$ relays, the RRS and PRS strategies are equivalent and optimal with respect to the e2e outage probability, provided that the decoding at relay $n$, $\forall n \in [1, N]$, is assumed to be successful if no outage event happens during the first transmission phase.
\end{theorem}

\begin{IEEEproof}
By definition, the e2e outage probability is the probability that the received SINR is smaller than a predefined threshold value. Let $\Gamma_{\mathrm{R}_{n_1}, \mathrm{D}}$ and $\Gamma_{\mathrm{R}_{n_2}, \mathrm{D}}$ denote the received SINRs at the destination $\mathrm{D}$ with respect to the RRS and PRS strategies, whereby $\mathrm{R}_{n_1}$ and $\mathrm{R}_{n_2}$ are the chosen best relays, respectively. Given the outage threshold $\gamma_\mathrm{th}$, in order to prove the outage equivalence of these two strategies, i.e. $\mathrm{Pr} \left\{\Gamma_{\mathrm{R}_{n_1}, \mathrm{D}} < \gamma_\mathrm{th}\right\} \Leftrightarrow \mathrm{Pr} \left\{\Gamma_{\mathrm{R}_{n_2}, \mathrm{D}} < \gamma_\mathrm{th}\right\}$ with the operator $\mathrm{Pr} \left\{ E \right\}$ being mathematical probability that event $E$ occurs, we need to demonstrate the following two statements.

\noindent {\bf Statement 1: $\boldsymbol{\mathrm{Pr} \left\{\Gamma_{\mathrm{R}_{n_1}, \mathrm{D}} < \gamma_\mathrm{th}\right\} \Rightarrow \mathrm{Pr} \left\{\Gamma_{\mathrm{R}_{n_2}, \mathrm{D}} < \gamma_\mathrm{th}\right\}}$.}

Regarding RRS strategy, for the relays in the decoding set, i.e. $r \in \mathcal{C}$ where $\mathcal{C}$ is assumed to be non-empty, it is obvious that $\min_{r \in \mathcal{C}} \Gamma_{\mathrm{S}, r} > \gamma_\mathrm{th}$ in the first transmission phase; otherwise, an outage event occurs and the relay must be out of $\mathcal{C}$. In the second transmission phase, assuming that relay $\mathrm{R}_{n_1}$ is chosen from the decoding set and serves as the best relay, i.e. $\Gamma_{\mathrm{R}_{n_1}, \mathrm{D}} =\max_{r \in \mathcal{C}} \Gamma_{r, \mathrm{D}}$; if $\Gamma_{\mathrm{R}_{n_1}, \mathrm{D}} < \gamma_\mathrm{th}$, then we immediately have $\max_{r \in \mathcal{C}}\min\left\{\Gamma_{\mathrm{S}, r}, \Gamma_{r, \mathrm{D}}\right\} < \gamma_\mathrm{th}$.

For the relays out of the decoding set, i.e. $r \in \overline{\mathcal{C}}$, it is clear that $\max_{r \in \overline{\mathcal{C}}} \Gamma_{\mathrm{S}, r} < \gamma_\mathrm{th}$ in the first transmission phase since they cannot decode the source signals correctly and, thus, $\Gamma_{r, \mathrm{D}}=0$, $\forall r \in \overline{\mathcal{C}}$ in the second transmission phase, which leads to $\max_{r \in \overline{\mathcal{C}}}\min\left\{{\Gamma_{\mathrm{S}, r}, \Gamma_{r, \mathrm{D}}}\right\} = 0 < \gamma_\mathrm{th}$.

Combining the above two cases implies that, if $\Gamma_{\mathrm{\mathrm{R}_{n_1}, \mathrm{D}}} < \gamma_{\mathrm{th}}$, then $\max_{r \in \{\mathcal{C}, \, \overline{\mathcal{C}}\}}\min\left\{\Gamma_{\mathrm{S}, r}, \Gamma_{r, \mathrm{D}}\right\} <  \gamma_{\mathrm{th}}$. On the other hand, according to the PRS criterion \eqref{Eq_PRS}, if $\max_{r \in \{\mathcal{C}, \, \overline{\mathcal{C}}\}}\min\left\{\Gamma_{\mathrm{S}, r}, \Gamma_{r, \mathrm{D}}\right\} <  \gamma_{\mathrm{th}}$, then we necessarily have $\Gamma_{\mathrm{\mathrm{R}_{n_2}, \mathrm{D}}} < \gamma_{\mathrm{th}}$. As a result, by using the transitive law, we conclude that $\Gamma_{\mathrm{\mathrm{R}_{n_1}, \mathrm{D}}} < \gamma_{\mathrm{th}}$ implies that $\Gamma_{\mathrm{\mathrm{R}_{n_2}, \mathrm{D}}} < \gamma_{\mathrm{th}}$, which confirms Statement 1.

An extreme case is that the decoding set $\mathcal{C}$ is empty, i.e. all of the relays are in outage. In such a case, for the RRS strategy, no relay will be active in the second transmission phase, which means that an outage event occurs at the destination. On the other hand, if the PRS is applied, although there is always one relay that can be chosen, no relay can successfully decode the received signals since the maximum of the SINRs at the first hop is smaller than the outage threshold $\gamma_\mathrm{th}$, which yields an outage event at the destination. Therefore, in such an extreme case, RRS and PRS have the same outage performance.

\noindent {\bf Statement  2: $\boldsymbol{\mathrm{Pr} \left\{\Gamma_{\mathrm{R}_{n_2}, \mathrm{D}} < \gamma_\mathrm{th}\right\} \Rightarrow \mathrm{Pr} \left\{\Gamma_{\mathrm{R}_{n_1}, \mathrm{D}} < \gamma_\mathrm{th}\right\}}$.}

As far as PRS strategy is concerned, assuming that relay $\mathrm{R}_{n_2}$ is chosen to be the best relay, if $\mathrm{R}_{n_2}$ can decode the received signals correctly, it means $\Gamma_{\mathrm{S}, \mathrm{R}_{n_2}} > \gamma_\mathrm{th}$. Accordingly, $\Gamma_{\mathrm{R}_{n_2}, \mathrm{D}} < \gamma_\mathrm{th}$ implies that $\max_{r \in \mathcal{C}} \Gamma_{r, \mathrm{D}} < \gamma_\mathrm{th}$. Now, according to the RRS criterion \eqref{Eq_RRS}, if $\max_{r \in \mathcal{C}} \Gamma_{r, \mathrm{D}} < \gamma_\mathrm{th}$, we have $\Gamma_{\mathrm{R}_{n_1}, \mathrm{D}} < \gamma_{\mathrm{th}}$. Therefore, by exploiting again the transitive law, we infer that, if $\Gamma_{\mathrm{R}_{n_2} , \mathrm{D}} < \gamma_\mathrm{th}$, we have $\Gamma_{\mathrm{R}_{n_1}, \mathrm{D}} < \gamma_{\mathrm{th}}$, which verifies Statement 2.

An extreme case occurs if the chosen relay $\mathrm{R}_{n_2}$ fails to decode the received signals, i.e. $\Gamma_{\mathrm{S}, \mathrm{R}_{n_2}} < \gamma_\mathrm{th}$. In this case, not matter whether $\Gamma_{\mathrm{S}, \mathrm{R}_{n_2}} < \Gamma_{\mathrm{R}_{n_2}, \mathrm{D}}$ or $\Gamma_{\mathrm{S}, \mathrm{R}_{n_2}} > \Gamma_{\mathrm{R}_{n_2}, \mathrm{D}}$, for the PRS strategy this means an outage event at the destination. On the other hand, even if RRS is applied in such a case, the outage event will surely happen as well. Hence, in such an extreme case, RRS and PRS perform identically.

Next, combining Statements 1 and 2 confirms that the RRS and PRS approaches are equivalent with respect to the e2e outage performance. Also, if a line-of-sight link exists between the source and the destination, the above equivalence can be readily proven by using the law of total probability. Finally, by recalling the fact that the RRS approach is outage-optimal in the sense that it is equivalent in outage behavior to the optimal DF relaying where all available relays are employed \cite[Theorem 1]{BletsasTWC0709}, we conclude that the PRS approach is outage-optimal as well.
\end{IEEEproof}

It is noteworthy that the above proof is general and independent of any prior assumptions, such as channel model, the number of antennas and their signal combining mechanisms, and the number of external interfering signals and their respective strengths.

Although RRS and PRS strategies are proven to be equivalent with respect to the e2e outage probability, their e2e symbol error rates are different, as specified below.

\begin{theorem}
The RRS relay selection approach has the same or lower e2e symbol error rate in comparison with the PRS.
\end{theorem}
\begin{IEEEproof}
For the RRS scheme, it always chooses the second-hop link with the highest SINR. For the PRS approach, however, the chosen link at the second hop is not always the one with the highest SINR among the available second-hop links. If the chosen second-hop link is exactly the one achieving the highest SINR, RRS and PRS attain the same e2e symbol error rate. On the other hand, if the chosen second-hop link does not achieve the highest SINR and does not introduce an outage event at the destination, it will surely yield a symbol error rate at the destination that is higher than the link with the highest SINR.
\end{IEEEproof}

Notice that, some other e2e performance measures such as average packet error rate and ergodic channel capacity behave similarly to the symbol error rate, since all of them are monotonically increasing with the e2e SINR.

Next, a case study is performed by explicitly deriving the CDFs of the e2e SINRs pertaining to the RRS and PRS strategies. In particular, both CCI and AWGN are taken into account at the relays and at the destination.

\begin{remark}[Effect of different criteria on successful decoding]
If the decoding set at the end of the first transmission is not determined by outage event but by the results of maximum likelihood decoding (MLD) performed at all relaying nodes,  the e2e outage performance of PRS will be worse than that of RRS, see e.g. \cite{AminTVT15}. The reason behind this assertion is that the cardinality of the decoding set determined by MLD should be larger than that determined by outage event, yielding lower e2e outage probably if RRS strategy is applied.
\end{remark}

\section{A Case Study: Cooperative DF Relaying with Co-Channel Interferences}
In this section, we explicitly derive the CDFs of the e2e SINR pertaining to the RRS and PRS strategies. For accuracy of description, we start with the signal model.

In the first transmission phase, the signal received at the relay in the $n^{\rm th}$ branch, is a combination of a faded noisy signal received from the source and faded CCI signals originating from $L$ external interferers. Mathematically, the signal received at relay $\mathrm{R}_{n}$ can be expressed as
\begin{equation}
\label{Eq_Rx-1}
y_{n}
= \sqrt {E_\mathrm{S}}\, \alpha _{\mathrm{S}, n} \, x_\mathrm{S} + \sqrt {E_\mathrm{I}} \sum\limits_{l = 1}^L {\beta _{l, n} } \, d_{l, n}  + w_n,
\end{equation}
where $E_\mathrm{S}$ denotes the energy of the transmit signal $x_\mathrm{S}$ at the source, with $\mathbb{E}\{|x_\mathrm{S}|^2\}=1$ ($\mathbb{E}$ is the expectation operator); $\alpha_{\mathrm{S}, n}$ is the channel coefficient from the source to relay $\mathrm{R}_n$, subject to Nakagami fading with parameter $m_h$; $d_{l, n}$ refers to the $l^{\rm th}$ interferer impacting the $n^{\rm th}$ relay, with $\mathbb{E}\{|d_{l, n}|^2\}=1$; the energy of each interferer is denoted by $E_\mathrm{I}$; $\beta _{l, n}$ is the channel coefficient from the $l^{\rm th}$ interferer to relay $\mathrm{R}_n$, subject to Nakagami fading with parameter $m_\mathrm{I}$; and the last term, $w_n$, denotes the AWGN at $\mathrm{R}_n$, with zero mean and variance $\sigma^2$.

\subsection{Reactive Relay Selection (RRS)}
\label{Subsec:RRS}
If a relay fails to decode the source message successfully, it remains silent and does not participate in the second transmission phase. In fact, only the relays belonging to the decoding set $\mathcal{C}$, consisting of the relays successfully decoding the source message, are allowed to be active and may forward the processed signal to the destination. That is, there are $k = |\mathcal{C}|$ intermediate relays which may forward their received signals to the destination. For each of these relays, $r \in \mathcal{C}$, the received signal at the destination can be written as
\begin{equation}
\label{Eq_Rx-2}
y_{r, \mathrm{D}}  = \sqrt {E_\mathrm{R}} \, \alpha _{r, \mathrm{D}} \, x_r  + \sqrt {E_\mathrm{I}} \sum\limits_{l = 1}^L {\beta _{l, \mathrm{D}} } \, d_{l, \mathrm{D}}  + w_\mathrm{D},
\end{equation}
where $E_\mathrm{R}$ denotes the energy of the signal $x_r$ transmitted from the $r^\mathrm{th}$ relay, with $\mathbb{E}\{|x_r|^2\}=1$; $\alpha_{r, \mathrm{D}}$ is the channel coefficient from the $r^{\rm th}$ relay to the destination, subject to Nakagami fading  with parameter $m_g$. Also, in (\ref{Eq_Rx-2}), the signal from the $l^{\rm th}$ interferer, $l = 1, 2, \cdots, L$, impacting the destination is represented by $d_{l, \mathrm{D}}$ with $\mathbb{E}\{|d_{l, \mathrm{D}}|^2\}=1$; the energy of the interfering signal is $E_\mathrm{I}$, as defined in \eqref{Eq_Rx-1}; the channel coefficient from the $l^{\rm th}$ interferer to the destination is given by $\beta _{l, \mathrm{D}}$, subject to Nakagami fading with parameter $m_\mathrm{I}$, and the last term, $w_\mathrm{D}$, denotes the AWGN at the destination, with zero mean and variance $\sigma^2$.

Based on \eqref{Eq_Rx-1}-\eqref{Eq_Rx-2}, in the sequel we derive the exact CDF of the e2e SINR, starting with its conditional CDF given the set of successfully decoding relays.

\newcounter{mytempeqncnt1}
\begin{figure*}[!t]
\setcounter{mytempeqncnt1}{\value{equation}}
\setcounter{equation}{8}
\begin{small}
\begin{equation} \label{Eq_CDF-Rn}
F_{\Gamma _{r, \mathrm{D}}}(\gamma)
 =  1-\sum_{i=0}^{m_g-1}\frac{1}{i! \, \Gamma(m_\mathrm{I} L)} \left(\frac{m_g \gamma}{\bar{\gamma}_h}\right)^{i} \left(\frac{m_\mathrm{I}}{\bar{\gamma}_\mathrm{I}}\right)^{m_\mathrm{I} L} \exp\left(-\frac{m_g \gamma}{\bar{\gamma}_g}\right)
\sum_{j=0}^i {j \choose k} \Gamma(k+m_\mathrm{I} L) \left(\frac{m_g \gamma}{\bar{\gamma}_g} + \frac{m_\mathrm{I}}{\bar{\gamma}_\mathrm{I}}\right)^{-(k+m_\mathrm{I} L)}.
\end{equation}
\end{small}
\setcounter{equation}{\value{mytempeqncnt1}}
\hrulefill
\vspace{-15pt}
\end{figure*}
\setcounter{equation}{4}

According to the RRS criterion given by \eqref{Eq_RRS}, given the decoding set $\mathcal{C}$ with cardinality $|\mathcal{C}| = k$, the e2e SINR can be readily shown as
\begin{equation}
\label{Eq_SINR-c}
\Gamma_{{\rm RRS} \vert k}
=  \max\limits_{r \in \mathcal{C}}{\Gamma_{r, \mathrm{D}}}
= \max\limits_{r \in \mathcal{C}}\frac{\gamma_{r, \mathrm{D}}}{1+\sum_{l=1}^{L}\gamma _{\mathrm{I}_{l, \mathrm{D}}}},
\end{equation}
where $\gamma _{r, \mathrm{D}}  = \left|\alpha _{r, \mathrm{D}}\right|^2 {E_\mathrm{R}/{\sigma^2}}$ denotes the received SNR at the destination with respect to the transmission from the $r^{\text{th}}$ relay, and $\gamma_{\mathrm{I}_{l, \mathrm{D}}}  = \left|\beta_{l, \mathrm{D}}\right|^2 {E_\mathrm{I}/{\sigma^2}}$ represents the interference-to-noise ratio (INR) at the destination with respect to the $l^{\rm th}$ interferer. Since the channel from the $r^\mathrm{th}$ relay to the destination is assumed to be subject to Nakagami fading with parameter $m_g$, the PDF of $\gamma_{r, \mathrm{D}}$ has Gamma distribution, given by
\begin{equation}
\label{Eq_GammaPDF}
f_{\gamma_{r, \mathrm{D}}}(\gamma)
=  \left(\frac{m_g}{\bar\gamma_g}\right)^{m_g}\frac{\gamma^{m_g-1}}{\Gamma(m_g)}\exp\left(-\gamma\frac{m_g}{\bar\gamma_g}\right),
\end{equation}
where $\bar\gamma_g \triangleq \mathbb{E}\left\{\left|{\alpha _{r, \mathrm{D}}}\right|^2\right\}E_\mathrm{R}/\sigma^2$ is the average SNR at the destination and $\Gamma(x) = \int_0^\infty{t^{x-1}e^{-t}\,\mathrm{d}t}$ is the Gamma function. In light of (\ref{Eq_GammaPDF}), the CDF of $\gamma_{r, \mathrm{D}}$ can be shown as
\begin{equation}
\label{Eq_GammaCDF}
F_{\gamma_{r, \mathrm{D}}}(\gamma)
=  1-\sum_{i=0}^{m_g-1}\frac{1}{i!}\left(\frac{m_g\gamma}{\bar\gamma_g}\right)^i\exp\left(-\frac{m_g\gamma}{\bar\gamma_g}\right).
\end{equation}

\indent Now, let $Y \triangleq \sum_{l = 1}^L {\gamma _{\mathrm{I}_{l,D} } }$. Since the variables $\gamma _{\mathrm{I}_{l, D}}$, $l=1, \cdots, L$, are independent and identically distributed (i.i.d.), the PDF of $Y$ can be readily expressed as
\begin{equation}
\label{Eq_Y}
f_Y \left( y \right)
=  \left({\frac{m_\mathrm{I}}{\bar \gamma _\mathrm{I}}}\right)^{m_\mathrm{I}L}
\frac{y^{m_\mathrm{I}L-1}}{\Gamma \left( {m_\mathrm{I}L} \right)} \, \exp\left(-\frac{m_\mathrm{I}}{\bar{\gamma}_\mathrm{I}}y\right),
\end{equation}
where $\bar\gamma_\mathrm{I} = \mathbb{E}\left\{\left| {\beta _{l,D} } \right|^2\right\} E_\mathrm{I}/\sigma^2$ denotes the average INR at the destination.

In light of \eqref{Eq_GammaPDF}-\eqref{Eq_Y}, the CDF of $\Gamma_{r, \mathrm{D}}$ shown in \eqref{Eq_SINR-c} can be readily computed as per \eqref{Eq_CDF-Rn} shown at the top of this page. Accordingly, the CDF of $\Gamma_{{\rm RRS} \vert k}$ given by \eqref{Eq_SINR-c} is given by
\setcounter{equation}{9}
\begin{equation}
F_{\Gamma_{{\rm RRS} \vert k}}(\gamma)
= F_{\Gamma _{r, \mathrm{D}}}^k(\gamma).
\end{equation}

Then, by recalling the law of total probability, the CDF of the e2e SINR can be computed as
\begin{small}
\begin{align}
F_{\Gamma_{\rm RRS}}(\gamma)
& =  \sum\limits_{k=0}^{N}{F_{\Gamma _{\mathrm{S}, n}}^{N-k}(\gamma) \, \left(1-F_{\Gamma _{\mathrm{S}, n}}(\gamma)\right)^{k} \, F_{\Gamma_{{\rm RRS} \vert k}}} \nonumber \\
& =  \sum\limits_{k=0}^{N}{F_{\Gamma _{\mathrm{S}, n}}^{N-k}(\gamma) \, \left[ \left(1-F_{\Gamma _{\mathrm{S}, n}}(\gamma)\right) \, F_{\Gamma _{r, \mathrm{D}}}(\gamma)\right]^k}  \nonumber\\
& =  \left[F_{\Gamma _{\mathrm{S}, n}}(\gamma) + \left(1-F_{\Gamma _{\mathrm{S}, n}}(\gamma)\right) \, F_{\Gamma _{r, \mathrm{D}}}(\gamma)\right]^N  \nonumber\\
& =  \left[F_{\Gamma _{\mathrm{S}, n}}(\gamma) + F_{\Gamma _{r, \mathrm{D}}}(\gamma) - F_{\Gamma _{\mathrm{S}, n}}(\gamma) F_{\Gamma _{r, \mathrm{D}}}(\gamma)\right]^N, \label{Eq_CDF_RRS}
\end{align}
\end{small}
\hspace{-3pt}where $\Gamma _{\mathrm{S}, n}$ denotes the received SINR at relay $n$ with respect to the signals transmitted from the source $\mathrm{S}$, and is defined similarly to $\Gamma_{r, \mathrm{D}}$ shown in \eqref{Eq_SINR-c}. Accordingly, the CDF of $\Gamma _{\mathrm{S}, n}$, i.e. $F_{\Gamma _{\mathrm{S}, n}}(\gamma)$, is similar to \eqref{Eq_CDF-Rn}, where the fading parameter $m_g$ and $\bar{m}_g$ should be replaced by $m_h$ and $\bar{m}_h$, respectively.

\subsection{Proactive Relay Selection (PRS)}
\label{Subsec:PRS}
By using a similar system model as aforementioned, if the PRS strategy given by \eqref{Eq_PRS} is applied, we derive the CDF of the e2e SINR at the destination. At first, according to \eqref{Eq_PRS}, we define an intermediate variable $T_n$ as
\begin{equation}
\label{Eq_Tn}
T_n \triangleq \min\left\{\Gamma_{\mathrm{S}, n}, \, \Gamma_{n, \mathrm{D}}\right\}, \quad n=1, \cdots, N
\end{equation}
where the CDF of $\Gamma_{n, \mathrm{D}}$ is identical to \eqref{Eq_CDF-Rn}.  Next, by recalling the theory of order statistics, the CDF of $T_n$ can be shown as
\begin{equation}
F_{T_n}(\gamma) = F_{\Gamma_{\mathrm{S}, n}}(\gamma) + F_{\Gamma_{n, \mathrm{D}}}(\gamma) - F_{\Gamma_{\mathrm{S}, n}}(\gamma)F_{\Gamma_{n, \mathrm{D}}}(\gamma).
\end{equation}
Subsequently, combining \eqref{Eq_Tn} with \eqref{Eq_PRS} and applying the theory of order statistics again, the CDF of the received SINR, $\Gamma_{\rm PRS}$, at the destination is given by
\begin{small}
\begin{equation}
 F_{\Gamma_{\rm PRS}}(\gamma)
 =  \left[F_{\Gamma _{\mathrm{S}, n}}(\gamma) + F_{\Gamma _{n, \mathrm{D}}}(\gamma) - F_{\Gamma _{\mathrm{S}, n}}(\gamma) F_{\Gamma _{n, \mathrm{D}}}(\gamma) \right]^N.  \label{Eq_CDF_PRS}
 \end{equation}
\end{small}
\hspace{-5pt}Since the CDFs, $F_{\Gamma _{n, \mathrm{D}}}(\gamma)$ in \eqref{Eq_CDF_PRS} and $F_{\Gamma _{r, \mathrm{D}}}(\gamma)$ in \eqref{Eq_CDF_RRS}, are completely the same as given by \eqref{Eq_CDF-Rn}, it is clear that \eqref{Eq_CDF_PRS} is identical to \eqref{Eq_CDF_RRS} and they yield the same e2e outage probabilities. This analytical comparison gives a convincing demonstration of Theorem~\ref{Theorem-1}.

Finally, we note that the Nakagami fading parameter $m_g$ in \eqref{Eq_GammaCDF} was assumed to take integer values. This assumption has only influence on the expression in \eqref{Eq_GammaCDF} and  \eqref{Eq_CDF-Rn}, but does not change the outage equivalence of the RRS and PRS strategies. Also, due to the rigorous mathematical proof as well as strict space limit, no simulation results are provided in the paper.

\section{Concluding Remarks}
Aiming at a deep understanding of two major relay selection strategies widely used in decode-and-forward relaying networking, namely, reactive relay selection (RRS) and proactive relay selection (PRS), this paper proved their equivalence and optimality with respect to the e2e outage probability, as well as the superiority of RRS over PRS with respect to the e2e symbol error rate. These fundamental relations between these relay selection approaches enable system designers to flexibly choose either of them as per the implementation preference.

\bibliographystyle{IEEEtran}
\bibliography{References}

\end{document}